
\magnification=1000
 \baselineskip=0.830truecm
\centerline {\bf OPEN QUESTIONS IN CLASSICAL GRAVITY}
\vskip 0.90truecm
\centerline {\bf Philip D. Mannheim}
\smallskip
\centerline {Department of Physics}
\centerline {University of Connecticut}
\centerline {Storrs, CT 06269-3046}
\centerline{mannheim@uconnvm.bitnet}
\vskip 0.50truecm
\noindent
{\it to appear in a special issue of Foundations of Physics
honoring  Professor Fritz Rohrlich on the occasion of his retirement,
L. P. Horwitz and A. van der Merwe Editors, Plenum Publishing Company, N.Y.,
Fall 1993.}
\vskip 1.00truecm
\centerline {\bf Abstract}
\vskip 0.50truecm
\noindent
We discuss some outstanding open questions regarding the validity and
uniqueness of the
standard second order Newton-Einstein classical gravitational theory. On the
observational side we discuss the degree to which the realm of validity of
Newton's Law
of Gravity can actually be extended to distances much larger than the solar
system
distance scales on which the law was originally established. On the theoretical
side we
identify some commonly accepted but actually still open to question assumptions
which go
into the formulating of the standard second order Einstein theory in the first
place.
In particular, we show that while the familiar second order Poisson
gravitational equation
(and accordingly its second order covariant Einstein generalization) may be
sufficient to
yield Newton's Law of Gravity they are not in fact necessary. The standard
theory
thus still awaits the identification of some principle which would then make it
necessary too. We show that current
observational information does not exclusively mandate the standard theory, and
that the
conformal invariant fourth order theory of gravity considered recently by
Mannheim and
Kazanas is also able to meet the constraints of data, and in fact to do so
without the
need for any so far unobserved non-luminous or dark matter.
\vskip 1.00truecm
\centerline {April, 1993~~~~~~~~~~~~~~~UCONN-93-1}
\vfill\eject
 \baselineskip=0.820 truecm
\hoffset=0.0truein
\hsize=6.5 truein
{\bf (1). Introduction}
\medskip
While a great deal of attention is currently being given to the formulating of
a quantum
theory of gravity, it is remarkable that no such similar attention is being
given to the
question of what the correct classical theory should be (despite the fact that
the
correct quantum theory would have to have the correct classical theory as its
classical
limit); and indeed, not only is the standard Newton-Einstein theory considered
to be the
correct classical theory, but the general view of the community is that the
whole issue
has long since been completely settled. Now while there is always some risk in
asserting
that a theory will be able to accommodate all future data no matter how
disparate, there
are of course some very good reasons to assign such status to the standard
Newton-Einstein
gravitational theory. Nonetheless, if a theory is to be the correct theory,
then not only
must it survive all future testing, but also it must have no open theoretical
loose ends.
In the present work we shall argue that the standard theory actually still has
some
unresolved open theoretical loose ends, and, further, that it may even be
facing a
critical observational moment of truth in regard to galactic rotation curve
data with
their apparent need for enormous amounts of so far unestablished non-luminous
or dark
matter. As regards these loose ends, we shall show that the standard theory is
only
sufficient to meet the demands of data and not in fact necessary; and indeed,
we shall
demonstrate this in detail by actually presenting an alternate theory of
gravity, a
fourth order conformal invariant one,$^{1-14}$ which also appears to be able to
handle the
available data, and to even do so without the need for any dark matter. Now
while the
interest of the present author is in the fourth order theory per se, as we
shall see, its
very study enables us to sharpen our evaluation and understanding of the
standard theory;
and if the standard theory is to continue to be regarded as the correct theory,
then it
would appear to us that the loose ends which we discuss below not only need to
be tidied
up, but that they need to be resolved in favor of the standard theory by the
identifying
of some currently unknown fundamental principle which would then ensure that
the
standard theory in fact be a necessary and not merely a sufficient theory of
gravity.
\medskip
{\bf (2). Observational status of the standard theory}
\medskip
Not only does most of our information regarding gravity derive from a study of
the solar
system, but also so does most of our intuition. In his study of the solar
system Newton
established a phenomenological Law of Gravity described by the gravitational
potential
$V(r)=-MG/r$. This potential exhibits three central aspects, first that the
potential
falls like $1/r$ at large distances, second that gravity is universally coupled
through
a fundamental constant $G$, and third that the gravitational potential is an
extensive
function of the amount of matter contained in the gravitational source. The law
was
originally established for weak gravity on distance scales from terrestrial up
to solar,
but since it possessed a universal aspect, it was intended that it would
continue to hold
on much larger distance scales where it had not been studied. Moreover, the law
actually
contains two separate and distinct kinds of universality, a universality in
dependence
on distance, and a universality in coupling strength. Thus we need to ask
whether the
potential really is $1/r$ on much larger distance scales, and whether universal
coupling
actually necessitates the introduction of the fundamental constant $G$ in the
first place.

As regards the fitting of the Newtonian potential to the solar system, the
theory was
challenged in two familiar though differing ways when some deviations were
found to the
Keplerian $v^2(r)=MG/r$ fall-off expected (Fig. (1)) of the rotational
velocities of the
planets as a function of their distance from the Sun. Firstly, the motion of
the planet
Uranus was found to not quite fit the Keplerian expectation, and so it was
proposed that
there exist yet another, heretofore undetected, outer solar system planet,
Neptune, which
would perturb Uranus in just the right way. Since it had not been observed at
the time it
was first proposed, Neptune constitutes an example of what is now called dark
matter; and
indeed, its subsequent detection was the first successful prediction of the
dark matter
idea. (For the moment its discovery is actually the only successful prediction
of dark
matter theory). As regards the inner solar system, it was found there that the
motion of
the planet Mercury was also not quite Keplerian, and with this planet being so
close, it
was very easy to ascertain that this time there were no previously unnoticed
nearby
gravitational sources. Hence the deviation of the motion of the planet Mercury
could not be
explained by dark matter. Rather, it required something far more radical, an in
principle
modification, with Einstein explaining the motion of Mercury by providing a
complete
reformulation of the entire theory of gravity through the introduction of
General
Relativity. Thus the general wisdom obtained from the solar system is that when
there is
an observational conflict, either change the matter distribution or change the
theory.

With the advent of galactic astronomy it became possible to repeat the solar
system type
study only on a much larger distance scale. In galaxies it is possible to
determine the
velocities of the stars themselves as they move in orbit around a galaxy (or,
equivalently, the velocities of the $HII$ emission regions which surround some
of the hot stars in the galaxies), as
well as the velocities of small amounts of neutral hydrogen gas which are also
typically
present. Thus one can study the ionized optical spectra directly ($HII$) or the
Lyman
alpha lines of the neutral gas ($HI$); and, indeed, the original $HII$ optical
studies of
Rubin, Ford, Thonnard and Burstein$^{15-18}$ yielded
rotation curves for spiral galaxies which not only showed no sign of any
Keplerian
fall-off, but which were even flat in structure out to the largest observed
distances.
However, eventually, after a concentrated effort to carefully measure the
surface
brightness (i.e. the stellar luminous matter distribution) of such galaxies, it
was
gradually realized (see e.g. Kalnajs,$^{19}$ and Kent$^{20}$) that the $HII$
curves
could essentially be described by a standard luminous Newtonian prediction
after all
(even in fact for galaxies such as UGC2885 for which the rotation$^{18}$ and
surface
brightness$^{20}$ data go out to as much as 80 kpc - see Fig. (2)), simply
because the
galaxies were extended sources rather than effective
pointlike ones such as the Sun, with the observed curves actually being, to a
very good
approximation, the luminous Newtonian expectation for such extended objects on
the
observed distance scales. Now, while the Newtonian
prediction for such galaxies would then actually show a fall-off at even larger
distances,
these distances were simply not explorable optically since the signal was too
weak; and,
generally, the optical data do not (because of their very nature) go out to a
large
enough distance to be able to explore any substantive deviations from luminous
Newtonian
behavior. In other words, it is only possible to observe $HII$ spectra in the
region
where the stars themselves are located,
i.e. within the extended luminous matter distribution region itself, a region
which turns
out to be Newtonian dominated. As we thus see, the flatness observed in the
optical data
has nothing to do with any asymptotic behaviour of the rotation curves, rather
it is
merely the Newtonian expectation in the interior optical disk region.

While the $HII$ data do not show any substantive non-canonical behaviour,
nonetheless,
the pioneering work of Rubin and coworkers brought the whole issue of galactic
rotation
curves into prominence and stimulated a great deal of study in the field. Now
it turns
out that the hydrogen gas is distributed in galaxies out to much farther
distances than
the stars, thus making the $HI$ studies ideal probes of the outer reaches of
the rotation
curves. One particularly prominent case is the galaxy NGC3198 which was studied
by
Bosma$^{21,22}$ and then by Begeman$^{23,24}$ (as part of the ongoing
University of
Groningen Westerbork Synthesis Radio Telescope survey) to interestingly again
show a flat rotation
curve, but this time in a region way outside the optical disk region charted in
Refs.
(25) and (26), to thus yield a flatness which was not compatible with the
Newtonian
fall-off expected in the outer region. (That $HI$ studies might lead to a
conflict with a
luminous Newtonian prediction was noted very early by Freeman$^{27}$ from an
analysis of
NGC300 and M33 and by Roberts and Whitehurst$^{28}$ from an analysis of M31).
Thus with the $HI$ studies (there are now about 30 well studied cases) it
became clear that there really was a problem with the interpretation of
galactic rotation
curve data, which the community immediately sought to explain by the
introduction of
galactic dark matter (the 'Neptune' type solution) since the Newton-Einstein
theory was
presumed to be universal. Fits to the $HI$ data have been obtained using dark
matter (Ref.
(26) provides a very complete analysis while Fig. (3) shows a fit to NGC3198
using the data of Refs. (24), (25) and (26)),
and while the fits are certainly phenomenologically acceptable, they
nonetheless possess
certain shortcomings. Far and away their most serious shortcoming is their ad
hoc nature,
with any found Newtonian shortfall then being retroactively fitted by an
appropriately
chosen dark matter distribution. In this sense dark matter is not a predictive
theory
at all but only a parametrization of the difference between observation and the
luminous
Newtonian expectation. As to possible dark matter distributions, none has
convincingly been derived from first principles  as a consequence of, say,
galactic
dynamics or formation theory (for a recent critical review see Ref. (29)),
with the most popular being the distribution associated with a modified
isothermal gas
sphere (a two-parameter spherical matter density distribution
$\rho(r)=\rho_o/(r^2+r_o^2)$
with an overall scale $\rho_o$ and an arbitrarily introduced non-zero core
radius $r_o$
which would cause dark matter to predominate in the outer rather than the inner
region -
even though a true isothermal sphere would have zero core radius). The appeal
of the
isothermal gas sphere is that it leads to an asymptotically logarithmic
galactic
potential and hence to asymptotically flat rotation curves, i.e. it is
motivated by the
very data that it is trying to explain. However, careful analysis of the
explicit dark
matter fits is instructive. Recalling that the inner region is already flat for
Newtonian
reasons, the dark matter parameters are then adjusted so as to join on to the
Newtonian
piece (hence the ad hoc core radius $r_o$) to give a continuously flat curve in
the
observed region. This matching of the luminous and dark matter pieces is for
the moment
completely fortuitous (van Albada and Sancisi$^{30}$ have even referred to it
as a
conspiracy) and not yet explained by galactic dynamics. Worse, in the fitted
region the
dark matter contribution is actually still rising, and thus is not yet taking
on its
asymptotic value. Hence the curves are made flat not by a flat dark matter
contribution
but rather by an interplay between a rising dark matter piece and a falling
Newtonian one, with the
asymptotically flat expectation not yet actually having even been tested.
(Prospects for
pushing the data out to farther distances are not good because the $HI$ surface
density
typically falls off exponentially fast at the edge of the explored region).
Beyond these
fitting questions (two dark matter parameters per galaxy is, however, still
fairly
economical),
the	outcome of the fitting is that galaxies are then 90$\%$ or so non-luminous.
Thus not
only is the Universe to be dominated by this so far undetected material, the
stars in
a galaxy are demoted to being only minor players, an afterthought as it were.
Since dark
matter only interacts gravitationally it is extremely difficult to detect (its
actual
detection with just the right flux would of course be a discovery of the first
magnitude),
and since it can be freely reparametrized as galactic data change or as new
data come on
line, it hardly qualifies as even being a falsifiable idea, the sine qua non
for a
physical theory. (While some possible dark matter candidate particle
may eventually be detected, the issue is not whether the particle exists at all
- it may
exist for some wholly unrelated reason, but rather whether its associated flux
is big
enough to dominate galaxies). Since the great appeal of Einstein gravity is its
elegance
and beauty, using such a band aid solution for it essentially defeats the whole
purpose, and is completely foreign to Einstein's own view of nature.

Given the results obtained from the solar system study, one can also consider a
galactic
'Mercury' type solution; and indeed, in the absence of revealed wisdom, it
would appear
that the community should actually have no choice but to explore both
viewpoints without
preconceived prejudice to see what may emerge, and only then to make a
judgement.
Nonetheless, this is not how the subject has been treated, with only a few
brave souls
being prepared to contemplate the possibility that the Newton-Einstein theory
may actually
require modification (how in principle it is possible to do this by exploiting
the
theory's loose ends while still retaining its tested features will be discussed
below in
Sec. (3)). From the point of view of non-relativistic theory alone, the motion
of a test
particle is given by inserting the law of force into Newton's second law of
motion.
There thus appear to be three possible empirical options. One is to retain
Newton's
gravitational force law  but then increase the number of sources (i.e. dark
matter); the
second is to change the law of force (e.g. Sanders$^{29}$ with his proposed
additional
exponential potential which is motivated by ideas based on a possible grand
unification
of the fundamental interactions); and the third is to change Newtonian dynamics
itself
(Milgrom$^{31}$). The Milgrom option in particular has been extensively
explored in the
literature and is currently phenomenologically viable (Ref. (32) provides a
recent
analysis). For our purposes in this paper however, we do not want to
address the non-relativistic galactic rotation curve question per se but rather
to
explore the freedom still available at the relativistic level, to see what
alternatives
relativistic theory might yield, and to only then go to the non-relativistic
limit
and attempt to fit data.

Motivated by a desire to have a theory of gravity which obeys the general
covariance
principle and also the additional demand of local conformal invariance (the
invariance
under arbitrary  local conformal stretchings
$g_{\mu \nu}(x) \rightarrow \Omega^2(x) g_{\mu \nu}(x)$ which is
now thought to be enjoyed by the other three fundamental interactions, the
strong,
electromagnetic and weak), Mannheim and Kazanas$^{1-14}$ have suggested that
gravity be
based not on the Einstein-Hilbert action
$$I_{EH}=-\int d^4x (-g)^{1/2}R^{\alpha}_{\phantom{\alpha} \alpha}/16 \pi G
\eqno(1)$$
but rather on the conformal invariant fourth order action
$$I_W = -\alpha \int d^4x (-g)^{1/2}
C_{\lambda\mu\nu\kappa}C^{\lambda\mu \nu\kappa} \eqno(2)$$
where $C_{\lambda\mu\nu\kappa}$ is the conformal Weyl tensor and $\alpha$ is a
purely
dimensionless coefficient. They explored the structure of this theory to
obtain$^{1}$ the
complete, exact, non-perturbative exterior vacuum solution associated with a
static,
spherically symmetric gravitational source such as a star in the theory, viz.
(up to
an unobservable overall conformal factor)
$$-g_{00}= 1/g_{rr}=1-\beta(2-3 \beta \gamma )/r - 3 \beta \gamma
+ \gamma r - kr^2 \eqno(3)$$
where $\beta$, $\gamma$, and $k$ are three appropriate integration constants.
As can be
seen, for small enough values of the linear and quadratic terms (i.e. on small
enough
distance scales) the solution reduces to the familiar Schwarzschild solution of
Einstein
gravity, with the conformal theory then enjoying the same successes as the
Einstein
theory on those distance scales. On larger distance scales the theory begins to
differ
from the Einstein theory through the linear potential term, and (with the
quadratic term
only possibly being important cosmologically, and with the $\beta\gamma$
product terms
being numerically negligible)
yields a non-relativistic gravitational potential
$$V(r)=-\beta/r+\gamma r/2 \eqno(4)$$
which can then be fitted to galactic data in the weak gravity limit.

In order to apply the potential of Eq. (4) to a galaxy, we note that spiral
galaxies
typically consist of an axisymmetric disk of stars with surface density
$N$exp$(-R/R_o)/2\pi R_o^2$ where $N$ is the number of stars and $R_o=1/\alpha$
is the
scale length of the luminous matter distribution. Integrating Eq. (4) over such
a
distribution yields$^{7}$ for the rotational velocities the extremely compact,
exact
expression
$$rV^{\prime}(r)=
(N\beta\alpha^3 r^2/2)[I_0(\alpha r/2)K_0(\alpha r/2)-
I_1(\alpha r/2)K_1(\alpha r/2)]$$
$$+(N\gamma\alpha r^2/2)I_1(\alpha r/2)K_1(\alpha r/2)
\eqno(5)$$
\noindent
an expression which behaves asymptotically as $N\beta /r+N\gamma r/ 2
-3N\gamma R_o^2/ 4 r $ as would be expected, and which is now ready for
fitting.

In a recent comprehensive analysis of the $HI$ rotation curves of spiral
galaxies
Casertano and van Gorkom$^{33}$ have found that the data fall into essentially
four general groups characterized by specific correlations between the maximum
rotation
velocity and the luminosity. In order of increasing luminosity the four groups
are
dwarf, intermediate, compact bright, and large bright galaxies. Thus as a first
attempt
at data fitting we have chosen to study one representative galaxy from each
group,
respectively the galaxies DDO154 (a gas dominated rather than star dominated
galaxy),
NGC3198, NGC2903, and NGC5907. This will immediately
enable us to test the flexibility of our theory, as well as confront the
systematics
apparent in dark matter fits to the same four groups where it is typically
found
that the more luminous the galaxy the proportionately less dark matter seems to
be needed.
For NGC3198 we use the rotation curve of Ref. (24) and the surface brightness
data of Refs. (25) and (26), for NGC 2903 the data are taken from Refs. (23)
and (25), for NGC 5907 from Refs. (30) and (34), and for DDO154 from Refs. (35)
and (36).
The fits we obtain are shown in Figs. (4-7) (the details are given in Ref. (7)
which also
assesses the quality level required of fitting - typically up to 5\% - or even
10\%
for gas dominated galaxies); and, as can be seen, the model does quite well
with
the data, being able to even reproduce the luminosity trend
found in the dark matter fits with the more luminous galaxies being
proportionately
more Newtonian. The mass to light ratios we find in the fits (viz.
$M/L$(154)=1.4,
$M/L$(3198)=4.2, $M/L$(2903)=3.5, $M/L$(5907)=6.1) are similar to those found
in the dark matter fits, with the mass to light ratio essentially increasing
with
luminosity to thus make the more luminous galaxies more Newtonian dominated.
For
the linear terms we obtain a galactic $1/\gamma=1/N\gamma_{star}$
of order the Hubble length (viz. $1/\gamma$(154)=4.0$\times$10$^{29}$cm.,
$1/\gamma$(3198)=2.9$\times$10$^{29}$cm., $1/\gamma$(2903)=1.3$\times$10$^{29}$
cm., $1/\gamma$(5907)=1.7$\times$10$^{29}$cm.), an intriguing fact which
suggests
a possible cosmological origin for $\gamma$. (Such a value for a galactic
$\gamma$ would then make $\gamma_{star}=\gamma/N$ completely
insignificant on solar system distance scales, to thus permit the Newtonian
potential
to nicely dominate there as required). As regards our fitting, we see that the
linear
term is essentially replacing the dark matter contribution (c.f. Figs. (3) and
(5)),
so that our fits appear to be able to handle
galactic rotation curves without the need for any dark matter at all. That our
model is
able to fit the data sample at all is perhaps initially surprising, since,
given its
linear potential, the model predicts that rotational velocities will ultimately
rise
asymptotically. However, because the galaxies are extended sources and not
pointlike ones,
we see that the falling Newtonian piece and the rising linear piece are able to
compensate each other to produce effectively flat rotation curves in the
observed region
(the right hand side of Eq. (5) is simply a very slowly varying function in
this region),
while also satisfying the luminosity correlation at the same time. Thus it
would appear
to us that at the present time one cannot categorically assert that the sole
gravitational
potential on all distance scales is the Newtonian one, and that, in the linear
potential,
the standard $1/r$ potential would not only appear to have a companion but to
have one
which would even dominate over it asymptotically. Indeed, the very need for
dark matter
in the standard theory may simply be due to trying to apply just the simple
Newtonian potential in a domain for which there is no prior (or even current
for that
matter)  justification. As regards our
fitting, we see that is not in fact necessary to demand flatness in the
asymptotic region
in order to obtain flat rotation curves in the explored intermediate region.
Thus, unlike
the dark matter fits, we do not need to know the structure of the data prior to
the
fitting, or need to adapt the model to a presupposed asymptotic flatness.
Further, not
only is our linear potential theory more motivated than the dark matter models
(Eq. (4)
arises in a fundamental, uniquely specified theory), it possesses one fewer
free
parameter per galaxy ($\gamma$ instead
of $\rho_o$ and $r_o$). Consequently, according to the usual criteria for
evaluating
rival theories, it is to be preferred. However, in order to fully merit this
status, the
theory has to  confront Einstein gravity head on, a task to which we now turn.
\bigskip
{\bf (3). The loose ends of the standard theory}
\bigskip
In his monumental development of General Relativity, Einstein set out to
accomplish three
basic goals; first, to extend to accelerating observers the relativistic
invariance
properties of physical systems he had established for observers moving with
uniform
velocity; second, to find a natural explanation for the observed equality of
inertial and gravitational masses; and third, to construct a field theory of
gravity, a
theory which at the time did not even obey Special Relativity.
While these issues are separate and distinct, the distinctions between them are
easily
blurred, leaving the impression that the only prescription that works is the
one based on
the use of the Einstein-Hilbert action of Eq. (1), with any possible departure
from this
prescription guaranteed to prove fatal - to both the candidate alternative and
to its
proponents. Nonetheless, it is worthwhile to go over the entire package piece
by piece,
to see whether there is still any place where it is possible to make changes.

As regards the issue of accelerating observers, we note first that this issue
has nothing
to do with gravity per se, since observers can accelerate in flat spacetime;
and, anyway,
gravity is not the only force that can produce accelerations, electromagnetism
can do so
too and that certainly is not thought to necessitate curvature. Thus the issue
for
accelerating observers in flat spacetime is how to construct equations of
motion on whose
physics
they all can agree. Consider, for instance, a standard free particle of
kinematic mass $m$
(the discussion here follows Ref. (6)) moving in flat spacetime according to
the
special relativistic generalization of Newton's second law of motion, viz.
$$m{d^2\xi^{\alpha} \over d \tau ^2} =0~~~,~~~R_{\mu \nu \sigma \tau}=0
\eqno(6) $$
where $d\tau=(-\eta_{\alpha \beta}d\xi^{\alpha} d\xi^{\beta})^{1/2}$ is
the proper time and $\eta_{\alpha \beta}$ is the flat spacetime metric, and
where we have
indicated explicitly that the Riemann tensor is (for the moment) zero.
Transforming to an arbitrary coordinate system $x^{\mu}$, and using the
definitions
$$\Gamma ^{\lambda}_{\mu \nu} = {\partial x^{\lambda} \over \partial
\xi^{\alpha}} {\partial ^2\xi^{\alpha} \over \partial x^{\mu} \partial
x^{\nu}}~~~~,~~~~g_{\mu \nu}= {\partial \xi^{\alpha} \over \partial
x^{\mu}} {\partial \xi^{\beta} \over \partial x^{\nu}}
\eta_{\alpha \beta} \eqno(7)$$
enables us to write Eq. (6) in the form
$$
m \left( {d^2x^{\lambda} \over d\tau^2} +\Gamma^{\lambda}_{\mu \nu}
{dx^{\mu} \over d\tau}{dx^{\nu } \over d\tau} \right) = 0~~~,
{}~~~R_{\mu \nu \sigma \tau}=0 \eqno(8)$$
with the invariant proper time then taking the form
$d\tau=(-g_{\mu \nu}dx^{\mu} dx^{\nu})^{1/2}$. As derived, Eq. (8) so far only
holds in
a strictly flat spacetime with zero Riemann curvature tensor, and indeed Eq.
(8) is only
a covariant rewriting of the special relativistic Newtonian second law of
motion,
i.e. it covariantly describes what an observer with a non-uniform
velocity in flat spacetime sees. (While the four velocity
$dx^{\lambda}/d\tau$ is a general covariant vector, its ordinary
derivative $d^2x^{\lambda}/d\tau^2$ (which samples adjacent points and
not merely the point where the four velocity itself is calculated) is
not, and it is only the entire left-hand side of Eq. (8) which transforms as a
general covariant four acceleration). Thus we see that in general it is Eq. (8)
which
should be taken as Newton's second law of motion (in flat spacetime) and not
Eq. (6),
to thus show that general covariance is not in principle related to curvature,
and that
it is a principle which should be imposed on all physical theories independent
of
whether or not the spacetime in which we live is curved.

Incidentally, we note in passing that Eq. (8) provides an answer to Mach's
criticism of
Newton's second law of motion. Mach sought to have some dynamical interplay
between local
and global physics which would select out (the non-relativistic limit) of Eq.
(6) as
special. Since Eq. (6) is not generally covariant, there is no need to have any
such
dynamics for it, rather it is Eq. (8) which is physically meaningful in
general, and for
a single particle Eq. (8) will reduce to Eq. (6) in flat spacetime in Cartesian
coordinates simply because it is covariantly allowed to do so. However, for a
system of
two particles in mutual
rotation around each other in flat spacetime it is impossible to find a
coordinate
transformation  which will bring Eq. (8) to the form of Eq. (6) for both of
them at once,
thus making Coriolis forces physical, again without the need for any mystical
interaction with the rest of the Universe. (That does not mean that dynamically
there is
no such interaction, only that it is not needed in order to meet Mach's
concerns). Thus
covariance itself appears to be the answer to Mach and not Einstein gravity per
se.
(Moreover, even when there is curvature, our ability to remove the Christoffel
symbols
over some local region (they can always be removed at one point because they do
not
transform as tensors) would, if anything, be enhanced by having little or no
curvature,
i.e. by having a weak coupling with the rest of the Universe rather than a
strong one).

Now, of course, we still do need to generalize Eq. (8) to include gravity since
gravity is also a physical force, and the
great insight of Einstein was to then realize that in a non-flat
spacetime if the gravitational field emerged as being in the Christoffel
symbol, then since the two terms on the left hand side of Eq. (8) have
the same coefficient, the equality of gravitational and inertial masses
would then be assured, with Eq. (8) being replaced by
$$ m \left( {d^2x^{\lambda} \over d\tau^2}
+\Gamma^{\lambda}_{\mu \nu}
{dx^{\mu} \over d\tau}{dx^{\nu } \over d\tau} \right)=0
{}~~~,~~~R_{\mu \nu \sigma \tau}\neq0,
\eqno(9)$$
\noindent
Equation (9) achieves two things, one is it establishes the metric as the
gravitational
field in the first place, and the other is it specifies how a test particle is
to couple
to gravity. Since geodesic motion follows from the covariant conservation of
the
energy-momentum tensor of test particles, and since this covariant conservation
occurs in
any theory which is coordinate invariant, the validity of Eq. (9) for test
particles in
no way entails the need to use the Einstein-Hilbert action. Thus the question
of what
(covariant) equation of motion is to fix the Christoffel symbols which are to
go into
Eq. (9) is in principle an independently explorable issue which is decoupled
from the
validity or otherwise of the equivalence principle, and constitutes the only
piece of the
entire Einstein package which would still appear to admit of further analysis.

As to the question of what specific equation of motion is in fact to be obeyed
by the
gravitational field, it is here that we run into the theoretical loose ends we
alluded
to in Sec. (1).  Thus while it can safely be said that gravity is indeed a
metric based
field theory, covariance alone is simply not sufficient to specify the
gravitational
action with more input being needed, since any covariant scalar action
whatsoever will
meet that requirement. Now in a sense it is quite remarkable that Einstein did
not appeal
to something like a fundamental principle which would then unambiguously
specify the
appropriate action, since that would be more in keeping with the viewpoint
which
motivated his explanation of the equality of gravitational and inertial masses.
Rather,
he chose the action of Eq. (1) simply because it would nicely recover the
standard
non-relativistic Newtonian phenomenology while also yielding calculable and
testable
covariant corrections to it, and it is to this aspect of the theory which we
now turn.

The variation of the action of Eq. (1) with respect to the metric yields, in
the
presence of a matter source with energy-momentum tensor $T_{\mu \nu}$,
the equation of motion
$$R_{\mu \nu} - g_{\mu \nu} R^{\alpha}_{\phantom {\alpha} \alpha}/2
= -8 \pi G T_{\mu \nu} \eqno(10)$$
\noindent
The great appeal of Eq. (10) is that, first, in the weak gravity limit it
reduces
to the second order Poisson equation
$$\nabla^2 \phi ({\bf r}) = g({\bf r}) \eqno (11)$$
with its familiar exterior Newtonian potential solution, viz.
$$\phi(r>R) = -{1 \over r} \int_{0}^{R} dr^{\prime} g(r^{\prime})
{r^{\prime}}^2  \eqno(12)$$
for a spherically symmetric, static source with radius $R$; while, second it
yields relativistic corrections to this non-relativistic theory. The
observational confirmation of these corrections on terrestrial to solar system
distance scales not only established the validity of the Einstein theory on
those scales but seems to have established it on all others too, even though
many other theories could potentially have the same leading perturbative
structure on a given
distance scale and yet differ radically elsewhere.

While Eq. (10) was chosen so as to yield Eq. (11), it is a curious but not
well appreciated fact$^{8}$ that there is actually not yet any evidence for the
independent validity of Eq. (11) in the interior of a gravitational source - in
the
exterior region alone one can only test the validity of the
(Newtonian) solution in that region, this not being sufficient to infer the
validity of
the Poisson equation itself in that and all other regions as well.
Specifically,
suppose a theory has a Newtonian potential as its exterior solution. Then for a
set of
such Newtonian sources we may determine the potential of the system via the
summation
$$\phi ({\bf r})=\sum {1 \over \vert {\bf r}-{\bf r}^{\prime} \vert} \eqno
(13)$$
a summation which of course reduces to Eq. (12) for an appropriately chosen
$g(r)$ which at this point would only need to be identified as the number
density
of such sources. Thus, when derived from the non-relativistic theory alone the
source needed for the Poisson equation is the number density, whereas, when
derived from the weak gravity limit of the Einstein theory, the source would be
the energy density. While these two densities are of course proportional in the
weak binding limit, we note that in general in order to recover Newton's
original
result that the gravitational potential be an extensive function of the amount
of
matter in the gravitating source, we only need establish proportionality to the
number density, and not necessarily to the energy density, a point on which we
will
capitalize below when we study the fourth order theory. Via Eqs. (12) and/or
(13) we
thus determine the Newtonian potential of a set of sources outside of the
region where
the sources themselves are located.

Now suppose we want to determine the potential inside a star. In the
non-relativistic
limit we would then take the individual atoms in the star to be the gravitating
sources
needed for the Poisson equation and calculate the
potential exterior to them but interior to the star again via Eqs. (12) and
(13),
i.e. again exterior to the (microscopic this time)
gravitating sources, to again give us an extensive potential in
the weak gravity and weak binding limits. Thus even interior to a star we are
still only using the exterior Newtonian potential in this limit. The only place
where
the Poisson
equation would have explicit independent consequences would be in the interior
of
the atoms, with a truly interior solution taking the form
$$\phi(r<R) = - {1 \over r} \int_{0}^{r} dr^{\prime} g(r^{\prime})
{r^{\prime}}^2 - \int_{r}^{R} dr^{\prime} g(r^{\prime}) r^{\prime} \eqno(14)$$
\noindent
However, interior to atoms Eq. (14) has never been tested, simply because
gravity
is not the predominant force there. Thus as of today the dynamical consequences
of the second order gravitational Poisson equation which go beyond those of
Newton's Law have yet to be tested, thus making it unnecessary to find a
covariant
generalization of the second order Poisson equation; rather one only needs to
recover its exterior solution in some limit.

As regards recovering this exterior solution, we recall of course that the
exterior solution to the Einstein equation of Eq. (10) is simply the familiar
Schwarzschild solution to the Ricci flat condition  $R_{\mu \nu}=0$, a solution
which nicely contains both the Newtonian potential and its relativistic
corrections;
and as we have just seen that is all we need for the standard non-relativistic
phenomenology which would just sum and count the number of such sources. Now
it was noted by Eddington in the very early days of Relativity that this same
exterior non-relativistic and relativistic phenomenology would follow if some
other function of the Ricci tensor vanished, either a higher power of it or
some
derivative of it instead, since such a situation would still admit of the
vanishing of the Ricci tensor itself. With this being precisely the case which
occurs in fourth order theories of gravity (theories which are just as
covariant
as the second order one), Eddington then threw out a challenge to the community
to tell him which one the correct theory of gravity was. This question has
generally
been ignored by the community and never satisfactorily resolved, even though it
goes to the very heart of the question of the uniqueness of gravitational
theory.

While the vanishing of a derivative or a higher power of the Ricci tensor
implies
the vanishing of the Ricci tensor, the reverse is not true, since the vanishing
of a function of the Ricci tensor could have other, possibly less desirable
solutions as well, and thus in order to answer Eddington's question it is
necessary to find such other solutions and identify their observational
implications. This was then the brief that Mannheim and Kazanas followed. The
variation of the conformal invariant action of Eq. (2) with respect to the
metric yields the equation of motion (see e.g. Ref. (1))
$$ W_{\mu \nu}={1\over2}g_{\mu\nu}(R^{\alpha}_{\phantom{\alpha}\alpha})
^{;\beta} _{\phantom{;\beta};\beta}
+ R_{\mu\nu\phantom{;\beta};\beta}^{\phantom{\mu\nu};\beta}
 -R_{\mu\phantom{\beta};\nu;\beta}^{\phantom{\mu}\beta}
-R_{\nu\phantom{\beta};\mu;\beta}^{\phantom{\nu}\beta}
 - 2R_{\mu\beta}R_{\nu}^{\phantom{\nu}\beta}
+{1\over2}g_{\mu\nu}R_{\alpha\beta}R^{\alpha\beta}$$

$$ -{2\over3} g_{\mu\nu}(R^{\alpha}_{\phantom{\alpha}\alpha})
^{;\beta}_{\phantom{;\beta};\beta}
+{2\over3}(R^{\alpha}_{\phantom{\alpha}\alpha})_{;\mu;\nu}
+{2\over3} R^{\alpha}_{\phantom{\alpha}\alpha} R_{\mu\nu}
-{1\over6}g_{\mu\nu}(R^{\alpha}_{\phantom{\alpha}\alpha})^2
=\left({1\over 4\alpha}\right)T_{\mu \nu} \eqno(15)$$
with $W_{\mu \nu}$ thus replacing the Einstein tensor in Eq. (10). Despite its
somewhat forbidding appearance, Mannheim and Kazanas$^{1}$ were able to solve
the
exterior static, spherically symmetric vacuum problem associated with Eq. (15)
exactly
to find the metric of Eq. (3) as its complete and exact solution. Thus the
Schwarzschild solution is nicely recovered as Eddington had already noted,
together with the new linear term whose potential role in galaxies was
discussed above.
Thus not only have we found the extra non-Schwarzschild solutions, we have seen
that rather than be undesirable, they may even enjoy some observational
support.

In the presence of a source, Mannheim and Kazanas went further to find$^{8}$
that for the associated interior problem Eq. (15) reduces without approximation
to the remarkably compact equation
$$\nabla^4 B(r) = (rB)^{\prime \prime \prime \prime}/r=f(r) \eqno(16)$$
where $B(r)=-g_{00}(r)$ and where the source is given by
$$f(r) = {3 (T^0_{{\phantom 0} 0} - T^r_{{\phantom r} r})/4
\alpha B(r)} \eqno(17)$$
\noindent
Thus the metric of Eq. (3) is immediately recognized as the (exact) solution to
a fourth order Laplace equation. In the presence of the source the fourth order
Poisson equation of Eq. (16) is also readily integratable to yield the exact
exterior solution (we ignore an uninteresting $w-kr^2$ particular integral)
$$B(r>R) =- {1 \over 6r} \int_{0}^{R} dr^{\prime} f(r^{\prime}) {r^{\prime}}^4
- {r \over 2} \int_{0}^{R} dr^{\prime} f(r^{\prime}) {r^{\prime}}^2
\eqno(18)$$
to thus both recover the linear and Newtonian potential terms of Eq. (3)
and enable us to express their coefficients as appropriate moments of the
source
function $f(r)$, viz.
$$ \beta(2-3\beta \gamma)
= {1 \over 6} \int_{0}^{R} dr^{\prime} f(r^{\prime}) {r^{\prime}}^4
{}~~~,~~~\gamma = -{1 \over 2} \int_{0}^{R} dr^{\prime} f(r^{\prime})
{r^{\prime}}^2 \eqno(19)$$
\noindent
We thus see that while a second order Poisson equation is sufficient to yield
Newton's Law of Gravity, it is not in fact necessary, with a fourth order
Poisson
equation being just as capable. While both the second and fourth order Poisson
equations contain the Newtonian potential as the dominant solution on shorter
distance
scales, our ability to use the second order Poisson equation on larger distance
scales
comes through the assertion that the potential is still strictly Newtonian on
those
distance scales also, something which while widely believed has yet to be
established.
Thus the case for the unique use of the second order Poisson equation (and
acordingly of
its second order Einstein generalization) on large distance scales has yet to
be
made; and of course, should such a case eventually actually be made, then after
all that,
dark matter would still have to be found.

While the expressing of the Newtonian coefficient as a fourth moment rather
than
a second moment integral of the energy-momentum tensor is unfamiliar, we should
note
that it violates no known observational information, since it is only required
to hold
within fundamental microscopic sources (or within macroscopic compact sources
for which
nothing at all is currently observationally known). As long as such
microscopic sources
put out $1/r$ potentials then the macroscopic weak gravity potential (the only
one which
has so far been explored observationally) still scales as their number in the
standard
way, this being all that is required for the standard Newtonian phenomenology,
as we
noted above. (In the Einstein theory the coefficient of the $1/r$ potential is
identified
as the second moment integral of the energy density of the source and called
its
mass. However, despite its commonplace familiarity, we are not
aware that such integrals have ever been performed for either microscopic or
macroscopic sources for that matter, and it is simply not known whether the
mass
of the Sun as determined by the gravitational motions of the planets is in fact
equal to the integral of its energy density, since the energy density function
of
the Sun is unknown). Finally, we also note that the integrals in Eq. (19) are
all proportional to the inverse of the fundamental coupling constant $\alpha$
introduced in the conformal action of Eq. (2). Thus our theory meets Newton's
final demand, namely that gravity be universally coupled; and interestingly we
see
that the theory is able to do so without the need to introduce $G$ at all.
Since
in the standard theory $G$ is not independently measurable anyway as it only
ever appears in the product $MG$ (its canonically quoted value just represents
the choice of units in which to measure mass), we see, that despite its
longevity,
Newton's constant may not actually play any fundamental role in physics at all.
This of course would be a severe problem for the Einstein-Hilbert action since
it gives fundamental status to $G$. (It would also be a severe problem for
string theory since the string tension is taken to be the Planck length - of
course string theories with their intrinsic fundamental string tension scale
are anyway already excluded by the underlying conformal
invariance assumption on which the action of Eq. (2) is based. We leave it up
to the
reader to decide whether the incompatibility of conformal gravity (a
four-dimensional field theory  which
is even renormalizable in its own right) and string theory is virtue or a
vice).

As the above argument shows, the standard second order theory is only
sufficient
to yield the Newtonian phenomenology and not necessary, with it being somewhat
dangerous to identify the complete covariant theory simply from a knowledge of
its first
few perturbative terms in a restricted kinematic regime. We believe this to be
a
major loophole in the standard theory which still needs to be addressed. Beyond
the
phenomenological issue of what relativistic theory is mandated by
non-relativistic
observation, the standard
theory comes with a few other loose ends. Specifically, since no principle was
offered which would uniquely select out the Einstein-Hilbert action (to thus
make the above discussion necessary as well as sufficient), no principle
was offered which would exclude a possible cosmological constant term, and
indeed
since its inception Einstein gravity has been plagued by the cosmological
constant
problem. The conformal theory makes three specific contributions to this
problem.$^{2}$ First, the starting conformal symmetry not only unambiguously
specifies the action (Eq. (2) is the unique conformally invariant action) but
by
doing so it thereby automatically excludes any cosmological constant term at
all
at the level of the input Lagrangian. Second, the
conformal theory possesses no Planck scale, so that any induced cosmological
term would not be required to be of Planck density. And third, if any
cosmological
term is induced in $T_{\mu\nu}$ as a consequence of the scale breaking Higgs
mechanism which is needed to induce masses into the conformal invariant theory
in the first place, then since the conformal $T_{\mu\nu}$ is traceless
(and remains so even after the symmetry breaking$^{6}$), any induced
cosmological term would have to be of the same order of magnitude as all the
other terms in $T_{\mu\nu}$ which for cosmology would mean the energy density
of the particles in the Universe. (This feature is exhibited explicitly in the
model
presented in Ref. (5) where
it is also shown that conformal cosmology has no flatness problem, another
major problem for the standard theory). Moreover, while it still needs to be
explored in detail, a cosmological constant term of order of magnitude the
cosmological energy density might provide a solution to the current age problem
of the Universe, with recent astrophysical measurements indicating an
age for the Universe which may be less than that of some of its constituents.
Thus it appears to us that the origin of the cosmological constant
problem in the standard theory may be traced back to an insufficiently
motivated choice of starting action, with the Einstein theory simply lacking an
additional invariance principle which would remove all ambiguity.

One final loose end in the standard theory is in our understanding not of the
gravitational side of the equation of motion but rather in our understanding
of the energy-momentum side. Unlike the situation that existed in Einstein's
time,
we no longer regard mass as being kinematical, but rather it is now thought of
as
being dynamically induced in theories in which there are no fundamental scales
at the level of the action at all, with all scales being induced dynamically in
the vacuum; and indeed this motivated Mannheim and Kazanas to study conformal
gravity in the first place. In conformal invariant theories the energy-momentum
tensor is kinematically traceless. Thus, if we believe that the microscopic
strong, electromagnetic and weak interactions are responsible for the masses of
elementary particles, and hence for the masses of macroscopic systems such as
stars, then we should naturally only consider traceless conformal invariant
energy-momentum tensors from the beginning. Consequently it is somewhat
difficult
to understand why any such energy-momentum tensor should be set equal to the
non-conformal invariant, non-traceless Einstein tensor as would be required by
Eq. (10).
Moreover, in the conformal case $T_{\mu\nu}$ could not be the familiar
non-traceless
kinematical test particle $T_{\mu\nu}$, since a conformal, dynamical particle
would
have to be accompanied by the Higgs field
which gave it its mass in the first place. Nonetheless, it turns out that such
dynamical particles still move on geodesics$^{6}$ with the Higgs field energy
and
momentum actually being found to simply decouple from their motion. Thus even
the standard
test particle Bianchi identity argument (another cornerstone of the standard
theory)
is irrelevant to geodesic motion if
particles have dynamical masses. As regards mass, we make one final comment.
The
equivalence principle relates gravitational and inertial masses. Since we now
believe that inertial masses are dynamical through the Higgs mechanism, we are
thus led to consider gravitational masses as being dynamical too. However, one
prominent gravitational mass, the Planck mass, is not dynamical. We are thus
led
naturally to consider removing it from physics; and thus the apparent lack of
compatibility between the Einstein theory and current ideas on mass and the
structure of the other fundamental interactions represents another loose end
for
the Einstein theory, one which could even hinder a possible unification of
gravity
with the other interactions.

We thus believe that we have identified some open questions for classical
gravity.
And even if the conformal theory advanced here were to fall by the wayside
(this
seems to be a somewhat widespread hope), nonetheless, that would not suddenly
make the standard theory right, and the questions we have asked of it would
still
require answering. Moreover, if the Einstein theory is in fact correct, then
these questions must not only be answered, but they must be resolved in its
favor, and,
indeed, a theory such as the Einstein theory deserves that they be answered. In
conclusion we would like to quote from Binney and Tremaine$^{37}$ who remarked
in their book (p. 637): "If a new theory of gravity is required, it will
ultimately be accepted because of its beauty and unifying properties rather
than
because it eliminates the need for dark matter". We believe that the conformal
theory presented here meets all of these requirements (since it possesses an
additional (conformal) symmetry beyond that of the Einstein theory, the theory
not only enjoys all the beauty and elegance of General Relativity, it even has
some more). Whether Binney and Tremaine are correct in saying that the theory
will ultimately be accepted remains however to be seen.

The author would like to acknowledge the ongoing help and encouragement of his
collaborator Demosthenes Kazanas. This work has been supported in part
by the Department of Energy under grant No. DE-FG02-92ER40716.00.
\noindent
\smallskip
\noindent
{\bf References}
\smallskip
\noindent (01). Mannheim, P. D., and Kazanas, D., Ap. J., 342, 635 (1989).
\smallskip
\noindent (02). Mannheim, P. D., Gen. Rel. Grav., 22, 289 (1990).
\smallskip
\noindent (03). Kazanas, D., and Mannheim, P. D., Ap. J. Suppl.
Ser., 76, 431 (1991).
\smallskip
\noindent (04). Mannheim, P. D., and Kazanas, D., Phys. Rev., D44, 417 (1991).
\smallskip
\noindent (05). Mannheim, P. D., Ap. J., 391, 429 (1992).
\smallskip
\noindent (06). Mannheim, P. D., Gen. Rel. Grav., 1993, in press.
\smallskip
\noindent (07). Mannheim, P. D., {\it Linear potentials and galactic rotation
curves},
UCONN-92-3, Ap. J., in press.
\smallskip
\noindent (08). Mannheim, P. D., and Kazanas, D., {\it Newtonian limit of
conformal
gravity and the lack of necessity of the second order Poisson equation},
UCONN-92-4.
\smallskip
\noindent (09). Mannheim, P. D., and Kazanas, D., {\it Exact vacuum solution to
fourth
order Weyl gravity}, in Proceedings of the Storrs meeting, the fourth
meeting (new series) of the Division of Particles and Fields of the
American Physical Society, University of Connecticut, August 1988.
Edited by K. Haller, D. C. Caldi, M. M. Islam, R. L. Mallett, P. D.
Mannheim, and M. S. Swanson, World Scientific Press, Singapore (1989).
\smallskip
\noindent (10). Mannheim, P. D., {\it Some exact solutions to conformal Weyl
gravity}, in
Nonlinear Problems in Relativity and Cosmology, Proceedings of the Sixth
Florida Workshop on Nonlinear Astronomy, University of Florida, October
1990. Edited by J. R. Buchler, S. L. Detweiler, and J. R. Ipser,
Annals of the New York Academy of Sciences, Vol. 631, 194 (1991).
\smallskip
\noindent (11). Kazanas, D., {\it Astrophysical aspects of Weyl gravity}, in
Nonlinear Problems in Relativity and Cosmology, Proceedings of the Sixth
Florida Workshop on Nonlinear Astronomy, University of Florida, October
1990. Edited by J. R. Buchler, S. L. Detweiler, and J. R. Ipser,
Annals of the New York Academy of Sciences, Vol. 631, 212 (1991).
\smallskip
\noindent (12). Mannheim, P. D., and Kazanas, D., {\it Current status of
conformal Weyl
gravity}, in Proceedings of the ``After the First Three Minutes"
Workshop, University of Maryland, October 1990. A. I. P. Conference
Proceedings No. 222, edited by S. S.
Holt, C. L. Bennett, and V. Trimble, A. I. P., N. Y. (1991).
\smallskip
\noindent (13). Kazanas, D., and Mannheim, P. D., {\it Dark matter or new
physics?}, in
Proceedings of the ``After the First Three Minutes" Workshop, University
of Maryland, October 1990. A. I. P. Conference Proceedings No. 222, edited by
S. S. Holt,
C. L. Bennett, and V. Trimble, A. I. P., N. Y. (1991).
\smallskip
\noindent (14). Mannheim, P. D., {\it Conformal gravity, cosmology and Newton's
law}, in
Proceedings of the XXth International Conference on Differential
Geometric Methods in Theoretical Physics, Baruch College/CUNY, New York,
June 1991. Edited by S. Catto and A. Rocha, World Scientific Press,
Singapore (1992).
\smallskip
\noindent (15). Rubin, V. C., Ford W. K., and Thonnard, N., Ap. J. (Letters),
225, L107 (1978).
\smallskip
\noindent (16). Rubin, V. C., Ford W. K., and Thonnard, N., Ap. J., 238, 471
(1980).
\smallskip
\noindent (17). Rubin, V. C., Ford W. K., Thonnard, N., and Burstein, D.,
Ap. J., 261, 439 (1982).
\smallskip
\noindent (18). Rubin, V. C., Burstein, D., Ford W. K., and Thonnard, N.,
Ap. J., 289, 81 (1985).
\smallskip
\noindent (19). Kalnajs, A. J., in Internal Kinematics and Dynamics of Disk
 Galaxies, IAU Symposium No. 100, ed. E. Athanassoula (Reidel, Dordrecht,
1983), p. 87.
\smallskip
\noindent (20). Kent, S. M., A. J. 91, 1301 (1986).
\smallskip
\noindent (21). Bosma, A., Ph. D. Thesis, Gronigen University, 1978.
\smallskip
\noindent (22). Bosma, A., A. J., 86, 1791 (1981).
\smallskip
\noindent (23). Begeman, K. G., Ph. D. Thesis, Gronigen University, 1987.
\smallskip
\noindent (24). Begeman, K. G., A. A., 223, 47 (1989).
\smallskip
\noindent (25). Wevers, B. M. H. R., van der Kruit, P. C., and Allen, R. J.,
A. A. Suppl. Ser., 66, 505 (1986).
\smallskip
\noindent (26). Kent, S. M., A. J. 93, 816 (1987).
\smallskip
\noindent (27). Freeman, K. C., Ap. J. 160, 811 (1970).
\smallskip
\noindent (28). Roberts, M. S., and Whitehurst, R. N., Ap. J. 201, 327 (1975).
\smallskip
\noindent (29). Sanders, R. H., A. A. Rev., 2, 1 (1990).
\smallskip
\noindent (30). van Albada, T. S., and Sancisi, R., Phil. Trans. R. Soc., A320,
447
(1986).
\smallskip
\noindent (31). Milgrom, M., Ap. J. 270, 365; 371, 384 (1983).
\smallskip
\noindent (32). Begeman, K. G., Broeils, A. H., and Sanders, R. H.,
Mon. Not. R. Astron. Soc. 249, 523 (1991).
\smallskip
\noindent (33). Casertano, S., and van Gorkom, J. H., A. J., 101, 1231 (1991).
\smallskip
\noindent (34). Barnaby, D., and Thronson, H. A., A. J. 103, 41 (1992).
\smallskip
\noindent (35). Carignan, C., and Freeman, K. C., Ap. J. (Letters) 332, 33
(1988).
\smallskip
\noindent (36). Carignan, C., and Beaulieu, S., Ap. J. 347, 760 (1989).
\smallskip
\noindent (37). Binney, J., and Tremaine, S., Galactic Dynamics, Princeton
University Press, Princeton, N.J. (1987).
\smallskip
\noindent
{\bf Figure Captions}
\smallskip
Figure (1). The Keplerian expectation for the velocities of the solar system
planets as
they orbit the Sun, with distance expressed in Astronomical Units (the mean
Earth - Sun
distance).
\smallskip
Figure (2). The luminous Newtonian expectation for the $HII$ velocity profile
of the galaxy UGC2885 obtained by integrating the Newtonian potential over
the stellar surface luminosity distribution (Ref. (20)) whose logarithm is
shown in the
upper diagram. The bars show the rotational velocity data of Ref. (18)
with their quoted errors. Note that all the data are confined to the 4
scale length stellar distribution region (a distribution which is parametrized
by a
leading exponential with $R_o=21.7$ kpc).
\smallskip
Figure (3). Phenomenological dark matter fit to the $HI$ rotational velocity
data of Ref.
(24) for the galaxy NGC3198. The luminous Newtonian contribution is obtained by
integrating the Newtonian potential over the stellar (Ref. (26)) and gaseous
(Ref. (25))
surface distributions shown in the upper two diagrams (the stellar is
represented via its
logarithm).
The dark matter contribution is obtained by integrating the Newtonian potential
over an
isothermal sphere. The full curve shows the overall prediction. Note that the
$HI$ data
extend to 11 stellar scale lengths, a distance which is way beyond the stellar
region.
\smallskip
Figure (4). The calculated rotational velocity curve associated with the
conformal
gravity potential $V(r)=-\beta/r+\gamma r/2$ for the representative gas
dominated dwarf irregular galaxy DDO154. The bars show the $HI$ rotational
velocity
data points of Ref. (36) with
their quoted errors, the full curve shows the overall theoretical velocity
prediction
as a function of distance from the galactic center, while the two indicated
dotted curves
show the rotation curves that the separate Newtonian and linear potentials
would produce
when integrated over the luminous matter distribution of the galaxy given in
Ref. (36).
No dark matter is assumed.
\smallskip
Figure (5). The calculated rotational velocity curve associated with the
conformal
gravity potential $V(r)=-\beta/r+\gamma r/2$ for the representative
intermediate sized
galaxy NCC3198. The bars show the $HI$ rotational velocity
data points of Ref. (24) with their quoted errors, the
full curve shows the overall theoretical velocity prediction as a function of
distance
from the galactic center, while the two indicated dotted curves show the
rotation curves
that the separate Newtonian and linear potentials would produce when integrated
over the
luminous matter distribution of the galaxy given in Refs. (25) and (26). No
dark matter
is assumed.
\smallskip
Figure (6). The calculated rotational velocity curve associated with the
conformal
gravity potential $V(r)=-\beta/r+\gamma r/2$ for the representative compact
bright
galaxy NGC2903. The bars show the $HI$ rotational velocity
data points of Ref. (23) with their quoted errors, the
full curve shows the overall theoretical velocity prediction as a function of
distance
from the galactic center, while the two indicated dotted curves show the
rotation curves
that the separate Newtonian and linear potentials would produce when integrated
over the
luminous matter distribution of the galaxy given in Ref. (25). No dark matter
is assumed.
\smallskip
Figure (7). The calculated rotational velocity curve associated with the
conformal
gravity potential $V(r)=-\beta/r+\gamma r/2$ for the representative large
bright
galaxy NGC5907. The bars show the $HI$ rotational velocity data points of Ref.
(30) with
their quoted errors, the full curve shows the overall theoretical velocity
prediction
as a function of distance from the galactic center, while the two indicated
dotted curves
show the rotation curves that the separate Newtonian and linear potentials
would produce
when integrated over the luminous matter distribution of the galaxy given in
Ref. (34).
No dark matter is assumed.
\end